\documentclass[pra,aps,twocolumn,superscriptaddress]{revtex4}
\usepackage{amsfonts}
\usepackage{amsmath}
\usepackage{amssymb}
\usepackage{graphicx}
\usepackage{color}
\usepackage[bookmarks=false,linkcolor=blue,urlcolor=blue,colorlinks,citecolor=blue]{hyperref}

\newcommand{\notop}{{{}}}

\begin{document}

\title{Nonlocal conductance spectroscopy of Andreev bound states:\\ Symmetry relations and BCS charges}
\author{Jeroen Danon}
\affiliation{Center for Quantum Spintronics, Department of Physics, Norwegian University of Science and Technology, NO-7491 Trondheim, Norway}
\author{Anna Birk Hellenes}
\affiliation{Center for Quantum Devices, Niels Bohr Institute, 2100 Copenhagen, Denmark}
\author{Esben Bork Hansen}
\affiliation{Center for Quantum Devices, Niels Bohr Institute, 2100 Copenhagen, Denmark}
\author{Lucas~Casparis}
\affiliation{Center for Quantum Devices, Niels Bohr Institute, 2100 Copenhagen, Denmark}
\affiliation{Microsoft Quantum -- Copenhagen, Niels Bohr Institute, University of Copenhagen, 2100 Copenhagen, Denmark}
\author{Andrew P. Higginbotham}
\affiliation{Center for Quantum Devices, Niels Bohr Institute, 2100 Copenhagen, Denmark}
\affiliation{Microsoft Quantum -- Copenhagen, Niels Bohr Institute, University of Copenhagen, 2100 Copenhagen, Denmark}
\author{Karsten Flensberg}
\affiliation{Center for Quantum Devices, Niels Bohr Institute, 2100 Copenhagen, Denmark}

\begin{abstract}
  Two-terminal conductance spectroscopy of superconducting devices is a common tool for probing
  Andreev and Majorana bound states. Here, we study theoretically a three-terminal setup, with two
  normal leads coupled to a grounded superconducting terminal. Using a single-electron scattering
  matrix, we derive the subgap conductance matrix for the normal leads and discuss its symmetries.
  In particular, we show that the local and the nonlocal elements of the conductance matrix have
  pairwise identical antisymmetric components. Moreover, we  find that the nonlocal elements are
  directly related to the local BCS charges of the bound states close to the normal probes and
  we show how the BCS charge of overlapping Majorana bound states can be extracted from experiments.
\end{abstract}
\maketitle

\label{sec:intro}

Tunneling spectroscopy is a well-established tool for studying normal metal-superconductor (NS) hybrid systems. In the context of topological superconductors, tunneling spectroscopy is widely used in attempts to identify Majorana bound states (MBSs) \cite{Mourik2012,Das2012,Deng2012,Deng2016,Nichele2017,Zhang2018,Vaitiekenas2018b}, the prediction being that a single isolated Majorana mode should yield a zero-bias peak that is quantized to a conductance of 2$e^2/h$ for temperatures much below the scale of the tunneling broadening \cite{Sengupta2001,Law2009,Flensberg2010}.
For overlapping Majorana states, the overlap gives distinct features in the two-probe conductance \cite{Flensberg2010,Hansen2016} which become very pronounced when probed with a quantum dot \cite{Leijnse2011,Deng2016,Prada2017,Clarke2017,Deng2018}.

While standard two-probe tunneling spectroscopy, with one normal and one grounded superconducting probe, can provide information about the subgap spectrum, it also has a severe limitation in that the interpretation of the data is ambiguous in the context of Majorana wires: Local Andreev states or so-called quasi-MBSs can give signatures that strongly resemble those of a truly topological zero mode \cite{Kell2012,Prada2012,Fleckenstein2018,Liu2017,Vuik2018}.
The reason for this is essentially that one local probe can, on general grounds, not confirm the true nonlocal nature of the MBS.

With the limitations of such a two-probe measurement, it is natural to investigate other types of finite-bias spectroscopy to access the nonlocal properties of the subgap states.
One approach is to use Coulomb-blockaded Majorana islands, where two normal probes are connected to the ends of the island and the proximitizing superconductor is floating~\cite{Albrecht2016,Albrecht2017,Vaitiekenas2018,Vaitiekenas2018b,Vaitiekenas2018c,Shen2018}.
Linear-response sequential transport is then possible only through states that have support at both ends of the island.
The spacing between the peaks in the zero-bias differential conductance as a function of a gate-induced potential offset provides information about the energy of the lowest-energy bound state on the island~\cite{VanHeck2016,Chiu2017}. Experiments have yielded results that could be consistent with the presence of overlapping exponentially localized MBSs~\cite{Albrecht2016,Shen2018,Vaitiekenas2018b} and also provided information about the quasiparticle dynamics on the island~\cite{Higginbotham2015,Albrecht2017}. However, numerical simulations showed that the observed detailed behavior of the bound-state energy could also indicate a significant contribution from trivial Andreev bound states~\cite{Chiu2017} and therefore the experimental observations cannot be regarded as conclusive evidence of the presence of MBSs on the islands.

\begin{figure}[t!]
\centerline{\includegraphics[width=0.8\columnwidth]{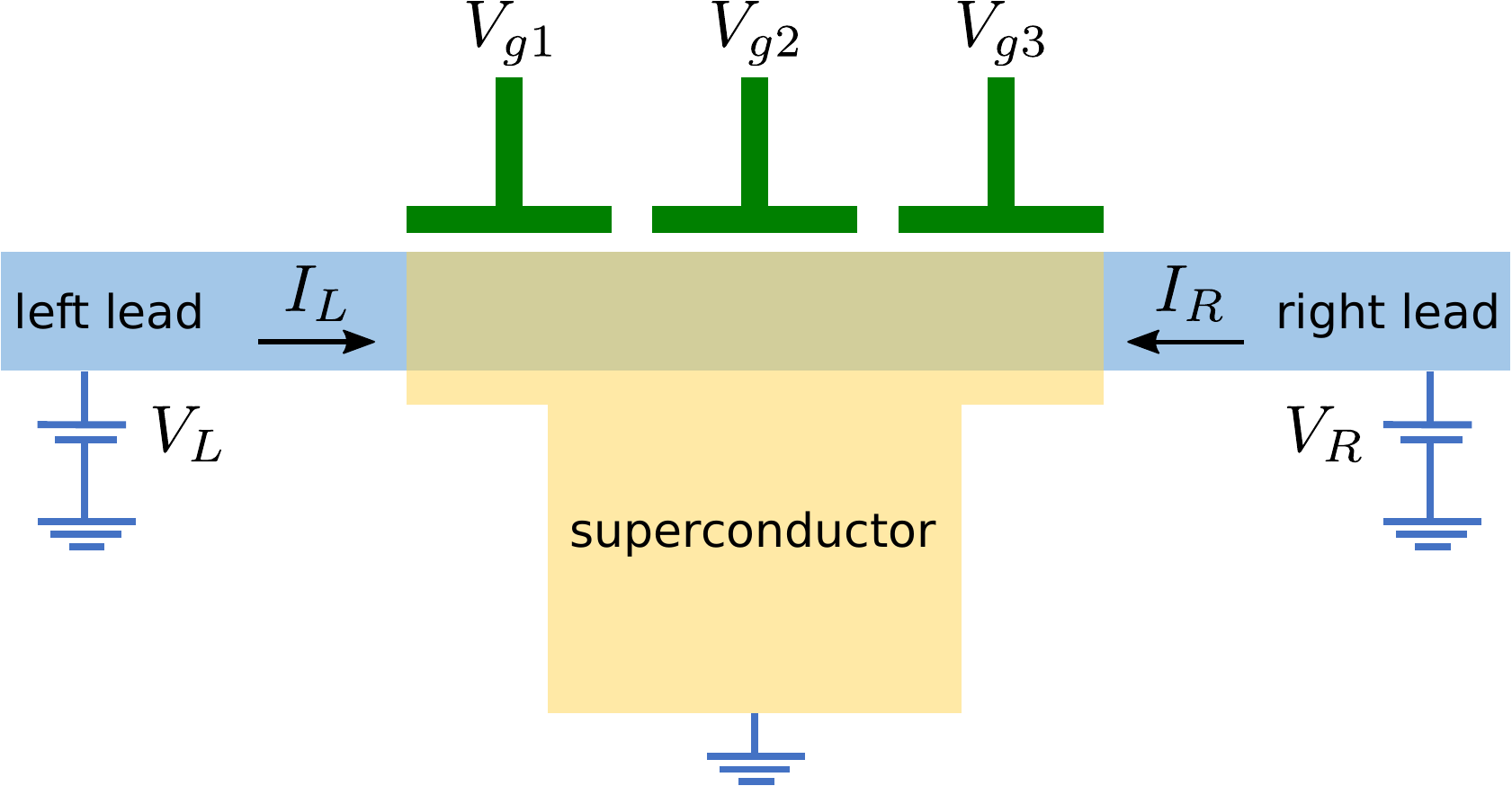}}
\caption{Layout of the device considered here: two normal leads are connected to a central grounded superconducting region, where the (local) potential can be controlled by electrostatic gates. Because no particles enter the superconductor at energies below the gap (instead, Andreev reflection can take place), the system is effectively a two-terminal device in terms of particle current.
\label{fig:layout}}
\end{figure}

Another approach is to consider a \textit{three-terminal} setup, with two normal local probes and a grounded superconducting probe (e.g., such as sketched in Fig.~\ref{fig:layout}), and a first step in this direction was already taken in \cite{Gramich2017}. Linear-response signals of three-terminal devices were also used in the context of the search for signatures of Cooper-pair splitting \cite{Hofstetter2009,Herrmann2010,Schindele2012}, following theoretical predictions \cite{Recher2001,Loss2000}.
Further, it was recently pointed out that for wires close to a topological transition, the nonlocal conductance gives information about the induced gap, the topological gap, as well as the coherence length \cite{Rosdahl2018}.

In this paper, we investigate this three-terminal setup in detail.
From quasiparticle-current conservation we derive a symmetry relation that dictates that the antisymmetric parts of the local and nonlocal conductance are equal for voltages below the gap.
Moreover, we show how the (experimentally accessible) nonlocal conductances contain detailed information about the electron and hole components of the bound states in the superconducting region, or more specifically, about the local BCS charge of the bound states, $|u(z)|^2-|v(z)|^2$, close to the two leads.
When compared with predictions from theoretical models, this could help to differentiate in practice between for example near-topological quasi-MBSs and trivial Andreev bound states.
In a parallel paper, these findings are investigated experimentally \cite{Menard2019}.

We first turn to the calculation of the current in the left and right normal leads, to which voltages $V_L$ and $V_R$ are applied, respectively.
Since we are interested in using the differential conductance to probe the subgap states, \textit{we will focus exclusively at voltages below the gap}: $V_{L,R}<\Delta$, where $\Delta$ is the gap in the grounded superconducting lead, see Fig.~\ref{fig:layout}.
This means that no quasiparticles are entering or leaving through the superconducting lead.

We start by using the conservation of probability current, corresponding to unitarity of the scattering matrix, to write the following identities (for more details, see App.~\ref{app:smatrix})
\begin{subequations}
	\label{conservations}
	\begin{align}\label{conservation11}
	&R_{\alpha }^{e}+A_{\alpha }^{e}+T_{\bar{\alpha}\alpha }^{e}+A_{\bar{\alpha}
		\alpha }^{e}=N_{\alpha },\\ \label{conservation22}
	&R_{\alpha }^{e}+A_{\alpha }^{h}+T_{\alpha\bar{\alpha} }^{e}+A_{\alpha\bar{\alpha} }^{h}=N_{\alpha },
	\end{align}
\end{subequations}
where $N_{\alpha }$ is the number of channels in lead $\alpha \in \{L,R\}$, $T_{\bar \alpha \alpha}^e$ is the total transmission probability of an electron from lead $\alpha$ to the opposite lead $\bar\alpha$, $A_{\bar\alpha {\alpha}}^e$ is the transmission probability of an electron in lead $\alpha$ to a \textit{hole} in lead $\bar\alpha$ (crossed Andreev reflection), and $R^e_\alpha$ and $A^e_\alpha$ denote the probability of reflection of an incoming electron in lead $\alpha$ as an electron or hole, respectively.
The first equation \eqref{conservation11} expresses the conservation of an incoming electron in lead $\alpha$, while the second equation \eqref{conservation22} expresses that an outgoing electron in lead $\alpha$ must have entered somewhere.

\begin{figure}[t!]
	\centerline{\includegraphics[width=1\columnwidth]{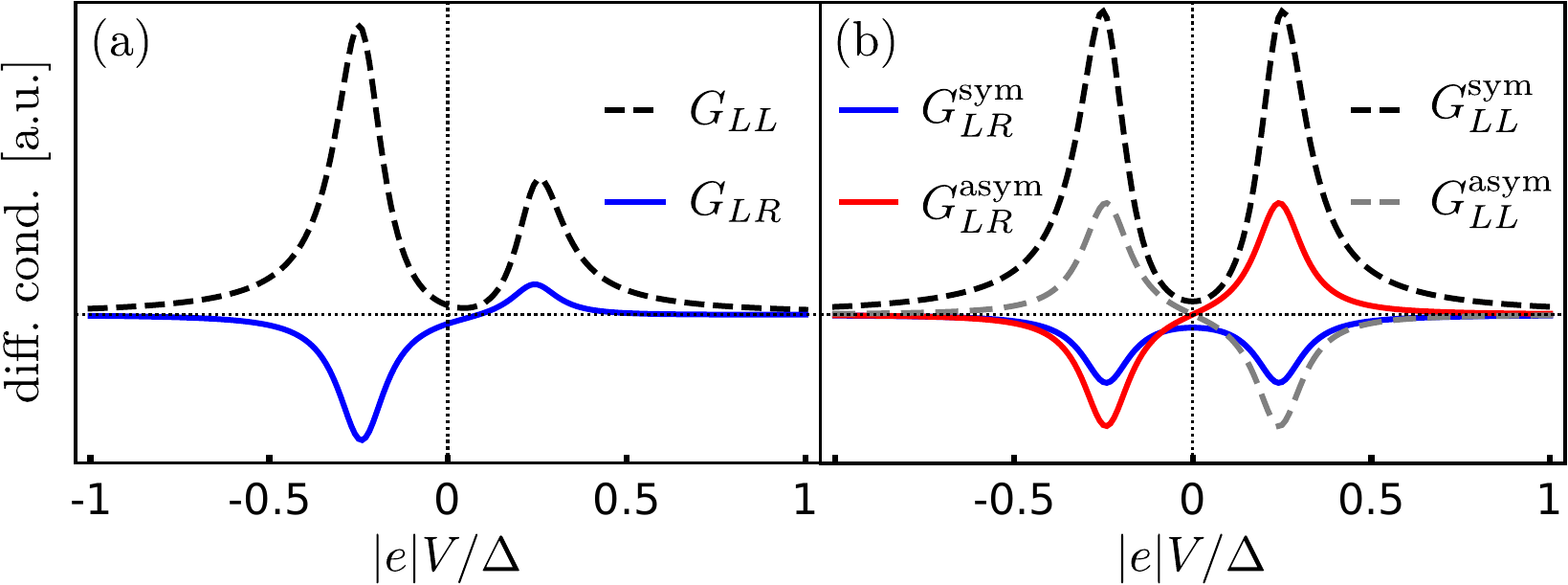}}
	\caption{(a) Local and nonlocal zero-temperature differential conductances for a single Andreev level bridging between the left and right leads, calculated using Eqs.~(\ref{current},\ref{eq:didv},\ref{eq:wm}) with $H=E_0\tau_3$. (b) The extracted symmetrized and antisymmetrized traces, demonstrating the symmetry relation pointed out in this paper $G_{LR}^{\rm asym}(V)=-G^{\rm asym}_{LL}(V)$. We used a bound-state energy $E_0=0.25\,\Delta$, set $\xi_{L,R}=0.024\,\Delta$ and $\gamma_{L,R}=0.04\,\Delta$ and chose all coherence factors $u_{L,R}$ and $v_{L,R}$ to be real and positive (see below for the exact definition of all parameters).
		\label{fig:symm}}
\end{figure}

Defining the positive direction of a current always to be \textit{into} the central scattering region, we can write \cite{Takane1992}
\begin{align}
I_{\alpha }= {} & {} -\frac{e}{h}\int_{-\infty }^{\infty }d\omega \, \tilde f_{\alpha}(\omega )
\left[ N_{\alpha }-R^e_{\alpha}(\omega )+A_{\alpha}^{e}(\omega )\right] \nonumber\\
{} & {} +\frac{e}{h}\int_{-\infty }^{\infty}d\omega \,\tilde f_{\bar{\alpha}}(\omega )
\left[ T_{\alpha \bar{\alpha}}^{e}(\omega )-A_{\alpha \bar{\alpha}}^e(\omega)\right],  \label{current}
\end{align}
where $\tilde f_{\alpha }(\omega )=f(\omega +eV_{\alpha }) - f(\omega)$, with $f(\omega )=1/(e^{\beta \omega}+1)$ the Fermi-Dirac distribution function (where $\beta = 1/k_{\rm B}T$ and $\omega$ is measured relative to the equilibrium chemical potential).
This equation then allows us to derive the elements of the differential-conductance matrix as
\begin{equation}
G_{\alpha \beta} = \frac{dI_\alpha}{dV_\beta},\label{eq:didv}
\end{equation}
where we will make the important assumption that all transmission and reflection coefficients do not depend on bias voltages, i.e., the voltages only enter through the distribution functions in the leads.
In that case, all elements $G_{\alpha\beta}$ only depend on one voltage (corresponding to the second index), so they have the form $G_{\alpha\beta}(V_\beta)$.

The second conservation law \eqref{conservation22} has the interesting consequence that the antisymmetric parts (in voltage) of $G_{\alpha\alpha}$ and $G_{\alpha\bar{\alpha}}$ are identical. This is easy to see when adding the two, e.g., for $\alpha=L$, and setting $V_L = V_R = V$
\begin{equation}\label{GLLmGLR}
G_{LL}(V)+G_{LR}(V)=- \frac{e^2}{h }\int_{-\infty
}^{\infty }d\omega\, f'(\omega+eV ) H(\omega),
\end{equation}
where $f'(\omega)$ is the derivative of the Fermi function
and
\begin{equation}\label{Hdef}
  H(\omega)=N_{L}-R^e_{L}(\omega )+A^e_{L}(\omega )-T_{LR}^e(\omega )+A_{LR}^e(\omega ).
\end{equation}
Now inserting $N_L-R^e_L(\omega)$ from Eq.~\eqref{conservation22}, we can obtain
\begin{equation}\label{GLLmGLR}
H(\omega)=A^e_{L}(\omega )+A^h_{L}(\omega )+A_{LR}^e(\omega )+A_{LR}^h(\omega ).
\end{equation}
Due to the general symmetry of the Andreev reflections $A^e(\omega)=A^h(-\omega)$, we then see that
\begin{equation}\label{GLLGLRasym}
G_{LL}(V)+G_{LR}(V)=G_{LL}(-V)+G_{LR}(-V).
\end{equation}
Defining the symmetric and antisymmetric components of the conductance as
\begin{align}
G^{\mathrm{sym}/\mathrm{asym}}_{\alpha\beta}(V)\equiv\frac{G_{\alpha\beta}(V)\pm G_{\alpha\beta}(-V)}{2},
\end{align}
we see that (\ref{GLLGLRasym}) implies that $G^{\rm asym}_{LL}(V) = -G^{\rm asym}_{LR}(V)$,
which is one of the main results of the paper, and is illustrated in Fig.~\ref{fig:symm}.
We emphasize that $G_{LL}(V)$ and $G_{LR}(V)$ are measured as functions of two \textit{different} voltages ($V_L$ and $V_R$, respectively); only when regarded as functions of the \textit{same} variable $V$ the relation \eqref{GLLGLRasym} holds.
Of course, a similar relation can be derived for $G_{RR}+G_{RL}$, and therefore, if one adds all four elements of the conductance matrix one gets a symmetric function, which is consistent with the fact that a two-terminal NS-junction has a symmetric differential conductance below the gap.

We now study the differential-conductance matrix in more detail, both its symmetric and antisymmetric parts.
We do this using a model where the central region has a single Andreev bound state that is coupled to the leads.
For metallic wide-bandwidth normal leads, we can use the expression \cite{AleinerReview} for the $S$-matrix,
\begin{equation}
\mathbf{S}=1-2\pi i\nu \mathbf{W}^{\dagger }\left( \omega -H+i\pi \nu \mathbf{WW}^{\dagger }\right) ^{-1}\mathbf{W},\label{eq:wm}
\end{equation}
where $\nu $ is the density of states in the leads.
The central superconducting region is described by a Bogoliubov-de Gennes Hamiltonian for a single level: $H=E_{0}\tau_z$, where  $\tau_{x,y,z}$ are the Pauli matrices in electron-hole space. The coupling matrix ${\bf W}$ follows as
\begin{equation}\label{eq:wmat}
{\bf W} =\left(
\begin{array}{cccc}
t_L u_{L} & -t_L v_{L}^{\ast} & t_R u_R & - t_R v_R^\ast \\
t_L v_{L} & -t_L u_{L}^{\ast} & t_R v_R & -t_R u_R^\ast
\end{array}
\right),
\end{equation}
where $t_\alpha u_\alpha (-t_\alpha v^*_\alpha)$ parameterizes the tunneling coupling between the electron(hole) component of the bound state and an electron state in lead $\alpha$.

The elements of the single-level $S$-matrix can now be inserted into the expressions for the conductance matrix, yielding explicit expressions for the zero-temperature differential-conductance matrix $G_{\alpha\beta}^0=(e^2/h) g_{\alpha\beta}^0$, from which finite-temperature expressions follow straightforwardly by convolution with $-f'(\omega)$.

The full results can be found in the App.~\ref{app:gmatrix}, and here we discuss only the limit where the scale of the energy of the bound state $E_0$ is much larger than its level broadening.
Focusing on the symmetric and antisymmetric components
we find close to the resonances, where $\omega \approx \pm E_0$,
\begin{align}
\label{gLRsymLimit}
g^{0,\mathrm{sym}}_{LR}(\omega)
& \approx -\xi_L \xi_R L^0(\omega),\\
g^{0,\mathrm{asym}}_{LR}(\omega)
& \approx -\xi_L \gamma_R  L^0(\omega)\, {\rm sign}(\omega),
\label{gLRasymLimit}\\
\label{gLLsymLimit}
g^{0,\mathrm{sym}}_{LL}(\omega)
& \approx [\gamma_L \gamma_R + \gamma_L^2 - \xi_L^2] L^0(\omega),\\
g^{0,\mathrm{asym}}_{LL}(\omega)
& \approx \xi_L \gamma_RL^0(\omega)\, {\rm sign}(\omega),
\label{gLLasymLimit}
\end{align}
where $\xi_\alpha = \pi\nu|t_\alpha|^2q_\alpha$ and $\gamma_\alpha = \pi\nu|t_\alpha|^2n_\alpha$ parameterize the coupling to the leads of the local BCS charges $q_\alpha = |u_\alpha|^2-|v_\alpha|^2$ and total (local) weights $n_\alpha = |u_\alpha|^2+|v_\alpha|^2$, respectively.
The function
\begin{equation}
L^0(\omega) = \frac{8E_0^2}{\left( E_0^2-\omega^2\right)^2 + 4\gamma^2E_0^2},\label{eq:L}
\end{equation}
where $\gamma = \gamma_L + \gamma_R$, is sharply peaked at $\omega = \pm E_0$, where $L^0(\pm E_0) = 2/\gamma^2$, and has a line width of $2\gamma$.

It is interesting to see that $g^{0,\text{sym}}_{LR}$ is proportional to the BCS charge at both terminals, $q_Lq_R$, while $g^{0,\text{asym}}_{LR}$ is proportional only to the BCS charge at the junction where the current is measured, in this case $q_L$ (this observation agrees with the rate-equation result derived in Ref.~\cite{Gramich2017}).
Therefore, the ratio of the peak heights
\begin{align}
Q_\alpha =
\frac{g^{0,\mathrm{sym}}_{\bar \alpha \alpha}(E_0)}{g^{0,\mathrm{asym}}_{\bar \alpha\alpha}(E_0)}\,{\rm sign}(E_0)
= \frac{q_\alpha}{n_\alpha},\label{eq:ratio}
\end{align}
provides a direct measure for the relative weight of the electron and hole components of the bound state close to end $\alpha$ of the wire.
This ratio is closely related to the parameter $\Lambda$ introduced in Ref.~\onlinecite{Hansen2018}, which can be extracted from the relative heights of neighboring conductance peaks in a Coulomb-blockaded setup~\cite{Albrecht2017,Shen2018} and reveals information about the actual wave function of the bound state.

Our results can easily be extended to include finite temperature, especially when $E_0 \gg k_{\rm B}T \gg \gamma_{L,R}$.
In that case one can approximate the expression given in (\ref{eq:L}) by $(2\pi/\gamma)\delta(E_0 - |\omega|)$, and convolution with $-f'(\omega)$ then straightforwardly yields
\begin{equation}
L(\omega,T) = \frac{\pi}{2 k_{\rm B}T \gamma} \,{\rm sech}^2\!\left( \frac{E_0 - |\omega|}{2k_BT} \right),
\end{equation}
which replaces $L^0(\omega)$ in the zero-temperature results.
We emphasize that finite temperature affects the line shape of all conductance peaks in the same way, and the relation found in (\ref{eq:ratio}) is thus valid at all temperatures, as long as all conductance peaks are well separated.

We thus found a general relation between the local BCS charge of a bound state close to the ends of the wire and the elements of the experimentally accessible differential-conductance matrix. In a sense, these results for an open (strongly coupled) setup complement those of Ref.~\onlinecite{Hansen2018} where the wire was treated as a Coulomb-blockaded island with a significant charging energy.

We now investigate the behavior of these BCS charges in more detail for the case where the scattering region is a quasi-one-dimensional semiconducting wire which has, besides proximity-induced superconductivity, strong spin-orbit coupling and a Zeeman splitting that can be made large enough to drive the wire into the topological regime.
We thus assume that we can describe the scattering region with the BdG Hamiltonian \cite{Oreg2010,Lutchyn2010}
\begin{equation}\label{OLwire}
{H} = \left(-\frac{\hbar^2\partial_z^2}{2m^*} - \mu - i\alpha\partial_z\sigma_y + V_{\rm Z} \sigma_z \right)\tau_z - \Delta\sigma_y\tau_y ,
\end{equation}
acting in the Nambu space $\{u_\uparrow(z), u_\downarrow(z), v_\uparrow(z), v_\downarrow(z)\}$,
where the Pauli matrices $\sigma$ act in spin space.
Here, $m^*$ is the effective electronic mass, $\mu$ the chemical potential in the wire, $\alpha$ characterizes the strength of the spin-orbit interaction, $\Delta$ is the proximity-induced pairing potential (assumed real for convenience), and $V_{\rm Z} = \frac{1}{2} g \mu_{\rm B}B$ corresponds to (half) the electronic Zeeman splitting in terms of the $g$-factor $g$ and the applied magnetic field $B$.

We immediately see that, quite generally, the local BCS charge in a bound state described by the Hamiltonian (\ref{OLwire}) can be related directly to the dependence of the bound-state energy $E_n$ on local potentials, which could be controlled through gates such as those sketched in green in Fig.~\ref{fig:layout}.
Indeed, if we add a term $V(z,a,b)\tau_z$ to the Hamiltonian to account for a local gate potential, where $V(z,a,b) = V_g$ for $a<z<b$ and zero otherwise, we obtain straightforwardly
\begin{equation}\label{dEdVx}
\frac{dE_n}{dV_g} = \int_a^b dz\, \big[ q_\uparrow(z)+q_\downarrow(z)\big],
\end{equation}
i.e., the slope of the bound-state energy as a function of $V_g$ corresponds to the total BCS charge locally at the position of the gate. Similarly, the dependence on a global gate voltage, which effectively controls $\mu$, is connected to the total integrated BCS charge.
We thus understand in very simple terms why the BCS charge, as deduced from the differential-conductance matrix, often changes sign at extrema of the conductance-versus-gate voltages traces, such as observed in \cite{Menard2019}. This of course assumes that the gate voltage only couples to the charge density and does not change other effective parameters such as the spin-orbit coupling strength.

Another common parameter to sweep in experiment is the magnetic field $B$. Assuming that the field predominantly affects the Zeeman splitting as in (\ref{OLwire}), i.e., neglecting any orbital contributions, we find
\begin{equation}\label{dEdZ}
\frac{dE_n}{dV_{\rm Z}} = \int dz\, \big[ q_\uparrow(z) - q_\downarrow(z) \big],
\end{equation}
which provides a connection between $E_n$ and the spin polarization of the BCS charge.
We note that in  the large-field limit, where the system is strongly spin-polarized, $dE_n/dV_{\rm Z}$ approaches the total BCS charge as well.

We now illustrate these findings with numerical examples.
First, we calculate the zero-temperature nonlocal conductance $g_{LR}^0$ using Eqs.~(\ref{current},\ref{eq:didv},\ref{eq:wm}) with a discretized version of the Majorana-wire Hamiltonian (\ref{OLwire}) to describe the scattering region.
We used $N=800$ lattice sites and we set the local coherence factors in (\ref{eq:wmat}) equal to the numerically found values for $u$ and $v$ at site 1 (for $L$) and site $N$ (for $R$).
We further used $\alpha = 0.28$~eV\AA, $\Delta=180~\mu$eV, $m^*=0.023\,m_e$, and we set the length of the wire to $L = 1500$~nm.
The resulting intersite hopping energy is $t = 471~$meV and we chose $\pi \nu |t_L|^2 = \pi \nu |t_R|^2 = t/2$.

\begin{figure}[t!]
\centerline{\includegraphics[width=1\columnwidth]{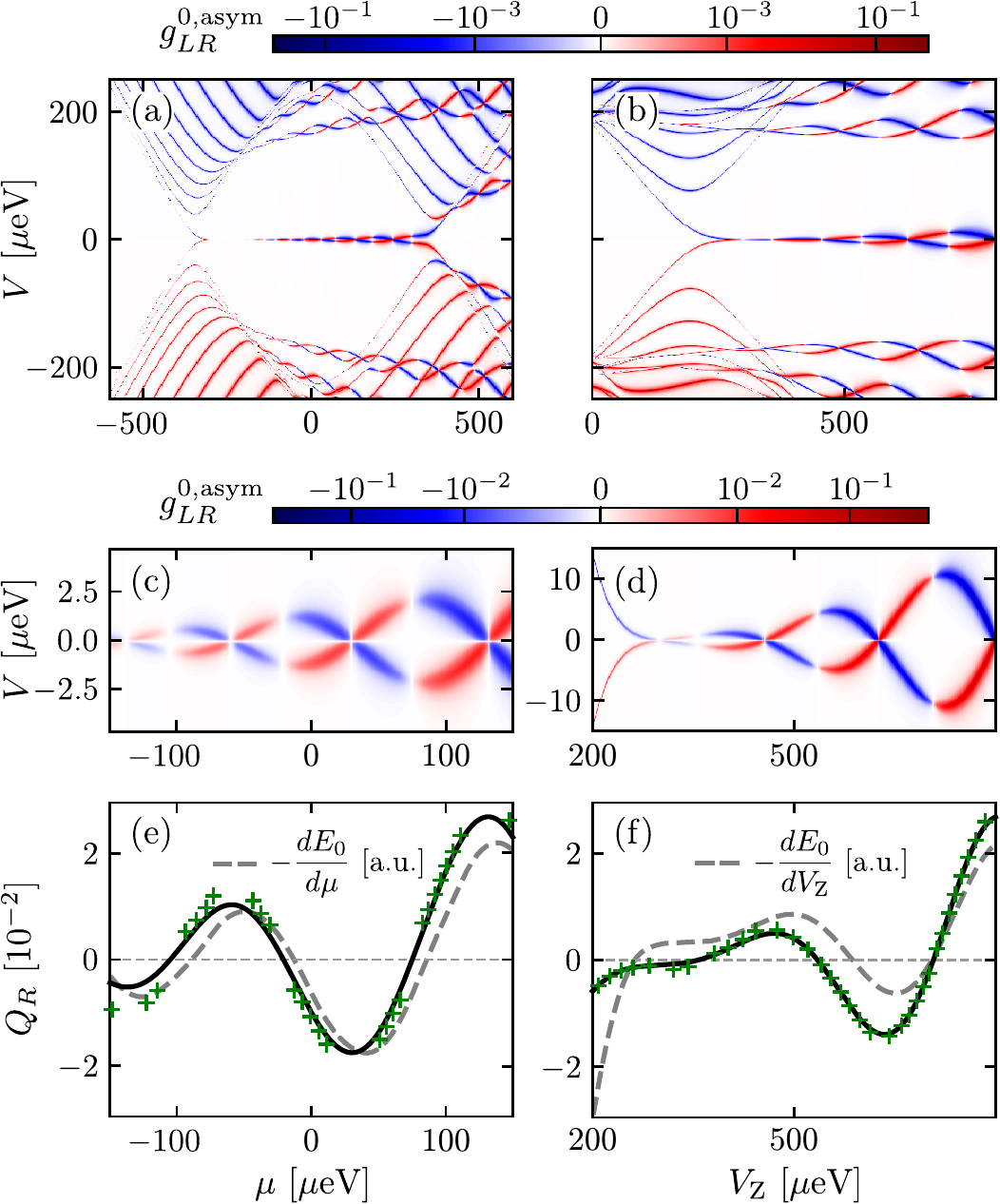}}
\caption{\label{fig:num1}
	(a--d) Numerically calculated antisymmetric component of the zero-temperature nonlocal conductance of a proximitized nanowire as a function of $\mu$ at $V_{\rm Z}=400~\mu$eV (a,c) and as a function of $V_{\rm Z}$ at $\mu=0$ (b,d).
	All other parameters are given in the text.
	Topological phase transitions occur in (a) at $\mu\approx \pm 350~\mu$eV and in (b) at $V_{\rm Z} \approx 200~\mu$eV.
	(c,d) Low-energy zooms inside the topological regime focusing on the contribution of the lowest-energy bound state only.
	(e,f) $Q_R$ of the lowest-energy state as it follows from Eq.~(\ref{eq:ratio}), using the calculated peak values of $g_{LR}^0$ (green crosses); $Q_R = q_R / n_R$ of the same state in the isolated wire, found from diagonalizing the Hamiltonian (\ref{OLwire}) (black solid lines); derivative of the bound-state energy, as found from diagonalizing the Hamiltonian (grey dashed lines).
}
\end{figure}

In Figs.~\ref{fig:num1}(a--d) we show the resulting antisymmetric component of the nonlocal conductance, both as a function of $\mu$ (a,c) and $V_{\rm Z}$ (b,d).
When we zoom in on the conductance associated with transport through the lowest-energy state (c,d), we already note a correlation between the magnitude and sign of $g_{LR}^{0,{\rm asym}}$ and the apparent slope of the energy of the bound state that is probed, as indeed predicted qualitatively by Eqs.~(\ref{gLRasymLimit},\ref{dEdVx},\ref{dEdZ}).

To investigate these relations in more detail, we show in Figs.~\ref{fig:num1}(e,f) the local BCS charge $Q_R$ as calculated with Eq.~(\ref{eq:ratio}) using the conductance peak values extracted from the numerical data (green crosses; we omitted regions where $g_{LR}^{0,{\rm asym}}$ is very small, leading to quick divergences due to small numerical inaccuracies).
We also calculate the ``actual'' BCS charges $q_R/n_R$ of the bound-state wave functions, in an unconnected wire, simply by diagonalizing the wire Hamiltonian (\ref{OLwire}), and we plot the resulting $Q_R$ (black solid lines).
There are no relative scaling factors involved, and the agreement with the $Q_R$ found from the conductance measurements is very clear.
For comparison, we also show the slope of the bound-state energy (in the unconnected wire) and, as expected based on Eqs.~(\ref{dEdVx},\ref{dEdZ}), we observe a stronger correlation with $Q_R$ in Fig.~\ref{fig:num1}(e) than in Fig.~\ref{fig:num1}(f), where the spin polarization of the bound state also plays a role.
The correlation in Fig.~\ref{fig:num1}(e) is of course also not expected to be perfect since the slope of the energy is related to the \textit{integrated} BCS charges whereas the conductance measurements probe the \textit{local} BCS charges at the ends of the wire.
In that sense, a comparison of $Q_R$ (as found from conductance measurements) and $dE_0/d\mu$ provides information about the degree of localization of the bound state close to the ends of the wire: a larger discrepancy implies more relative weight of the bound state in the center of the wire.
This could be a valuable tool in the search for and characterization of (quasi-)MBSs.

In conclusion, we have studied a three-terminal device with one of the terminals being a superconducting lead and the two other normal leads. From the general scattering matrix of this system (for quadratic Hamiltonians), we showed that there is a correspondence between the antisymmetric local and nonlocal differential conductances below the superconducting gap. For a single (Andreev bound state) level in the central region there is a furthermore a correspondence between nonlocal conductance and the BCS charges of the bound state (at the terminals). This allows for a study of the electron-hole texture of in-gap bound states and, in particular, it offers a way to test for the signatures characteristic for Majorana bound states with a small overlap, namely that the BCS charge and the energy splitting oscillates out of phase \cite{Ben-Shach2015,Hansen2018}.

\acknowledgments We acknowledge support by the Danish National Research Foundation, the
Deutsche Forschungsgemeinschaft (DFG, German Research Foundation)---Projektnummer 277101999---TRR 183 (project C01), as well as the Research Council of Norway through its Centers of Excellence funding scheme, project number 262633, QuSpin.

\appendix

\section{Symmetries of the scattering matrix}
\label{app:smatrix}

In this section we discuss the properties of the scattering matrix for energies below the superconducting gap.
We consider a system with two normal leads and one superconducting lead with a gap $\Delta$
For energies below $\Delta$, no single particle states can propagate into the gapped superconductor and in that regime the $S$-matrix can thus be written as a two-terminal $S$-matrix.
Naturally, if the superconductor is gapless this is no longer true.

Thus, assuming a gapped superconductor, we write the $S$-matrix for $\omega<\Delta$ as
\begin{equation}
\mathbf{S}(\omega)=\left(
\begin{array}{cc}
\mathbf{r}_{L}^{{}}(\omega) & \mathbf{t}_{LR}^{{}}(\omega) \\
\mathbf{t}_{RL}^{{}}(\omega) & \mathbf{r}_{R}^{{}}(\omega)%
\end{array}%
\right) ,
\end{equation}
where the reflection matrix and transmission matrices have a 2$\times$2 block structure in particle-hole space,
\begin{align}
\mathbf{r}_\alpha = {} & {} \left(
\begin{array}{cc}
r_{ee,\alpha} & r_{eh,\alpha} \\
r_{he,\alpha} & r_{hh,\alpha}%
\end{array}%
\right),\\
\mathbf{t}_{\bar \alpha \alpha} = {} & {} \left(
\begin{array}{cc}
t_{ee,\bar \alpha \alpha} & t_{eh,\bar \alpha \alpha} \\
t_{he,\bar \alpha \alpha} & t_{hh,\bar \alpha \alpha}%
\end{array}%
\right) .  \label{rtdef}
\end{align}%
Here $r_{eh,\alpha}$ is the usual Andreev reflection matrix for lead $\alpha$ and $t_{eh,\bar\alpha\alpha}$ is the matrix describing cross-Andreev reflection (CAR) from lead $\alpha$ to $\bar \alpha$.
All four components of ${\bf r}_\alpha$ are $N_\alpha \times N_\alpha$ matrices, where $N_{\alpha }$ is the number of channels in lead $\alpha$, and the components of ${\bf t}_{\bar\alpha\alpha}$ are thus $N_{\bar\alpha}\times N_{\alpha}$ matrices.

From the unitary of the $S$-matrix, $S^\dag S=SS^\dag=1$ we obtain that
\begin{subequations}
	\begin{align}
	\label{conservation1}&\mathbf{r}_{\alpha }^{\dag }\mathbf{r}_{\alpha }^{{}}+\mathbf{t}_{\bar{\alpha}\alpha }^{\dag }\mathbf{t}_{\bar{\alpha}\alpha }^{{}}=1,\\
	&\label{conservation2} \mathbf{r}
	_{\alpha }^{{}}\mathbf{r}_{\alpha }^{\dag }+\mathbf{t}_{\alpha\bar{\alpha}}^{{}}
\mathbf{t}_{\alpha\bar{\alpha}}^{\dag }=1.
	\end{align}
\end{subequations}
Both equations are consequences of conservation of probability current.
The first equation, (\ref{conservation1}), expresses conservation of probability current of a particle (first diagonal element) or a hole (second diagonal element) from lead $\alpha$.
The second equation, (\ref{conservation2}), expresses that the total probability current entering lead $\alpha$ must originate from somewhere.
In terms of the total reflection and transmission probabilities
\begin{equation}
\begin{aligned}
R_{\alpha }^{e}={} & {} \mathrm{Tr}_\alpha^\notop[r_{ee,\alpha }^{\dagger }r_{ee,\alpha
}^{{}}],
& T_{\bar{\alpha}\alpha }^{e}= {} & {} \mathrm{Tr}_\alpha^\notop[t_{ee,\bar{\alpha}\alpha }^{\dagger }t_{ee,\bar{\alpha}\alpha }^{{}}],\\
A_{\alpha }^{e}= {} & {} \mathrm{Tr}_\alpha^\notop[r_{he,\alpha }^{\dagger
}r_{he,\alpha }^{{}}],
& A_{\bar{\alpha}\alpha }^{e}= {} & {} \mathrm{Tr}_\alpha^\notop[t_{he,\bar{\alpha}\alpha }^{\dagger
}t_{he,\bar{\alpha}\alpha }^{{}}],  \label{RATedef}
\end{aligned}
\end{equation}
we directly obtain obtain Eqs.~(1) in the main text, as well as two similar conservation equations for holes.

The $S$-matrix also obeys electron-hole symmetry
\begin{equation}\label{PSP}
S(\omega )= \mathcal{P}S(-\omega )\mathcal{P}^{-1},
\end{equation}
where $\mathcal{P}$ is the particle-hole symmetry operator which is an antiunitary operator and reads explicitly $\mathcal{P}=\tau_x \mathcal{K}$, where the $\tau_i$ (with $i=x,y,z$) are Pauli matrices operating in electron-hole space and $\mathcal{K}$ is the complex conjugation operator.
In this basis, Eq.~\eqref{PSP} leads to
\begin{equation}
\begin{aligned}
r_{ee}^{{}}(\omega )= {} & {} r_{hh}^{\ast }(-\omega ) ,
& r_{eh}^{{}}(\omega)= {} & {} r_{he}^{\ast }(-\omega ) ,\\
t_{ee}^{{}}(\omega )= {} & {} t_{hh}^{\ast }(-\omega) ,
& t_{eh}^{{}}(\omega )= {} & {} t_{he}^{\ast }(-\omega ) .  \label{PHsym}
\end{aligned}
\end{equation}
These symmetries yield the following relations between the total reflection and transmission coefficients:
\begin{equation}
\begin{aligned}
R_{\alpha }^{e}(\omega)= {} & {} R_{\alpha }^{h}(-\omega),
& A_{\alpha }^{e}(\omega)= {} & {} A_{\alpha }^{h}(-\omega),\\
T_{\bar{\alpha}\alpha }^{e}(\omega)= {} & {} T_{\bar{\alpha}\alpha }^{h}(-\omega),
& A_{\bar{\alpha}\alpha }^{e}(\omega)= {} & {} A_{\bar{\alpha}\alpha }^{h}(-\omega).\label{RATsym}
\end{aligned}
\end{equation}

\section{The differential-conductance matrix}
\label{app:gmatrix}

As explained in the main text, we define the differential-conductance matrix as
\begin{equation}
{\bf G} =
\left( \begin{array}{cc} G_{LL} & G_{LR} \\ G_{RL} &
G_{RR}\end{array}\right) \equiv \left( \begin{array}{cc}
\frac{dI_{L}}{dV_{L}} & \frac{dI_{L}}{dV_{R}} \\ \frac{dI_{R}}{dV_{L}} &
\frac{dI_{R}}{dV_{R}}\end{array}\right).\label{diffcondmatrix1}
\end{equation}
Inserting the expression for the currents $I_\alpha$ as given by Eq.~(2) in the main text and taking all derivatives yields
\onecolumngrid
%\begin{widetext}
\begin{equation}
{\bf G}
=\frac{e^2}{h }\int d\omega \left( \begin{array}{cc} h(\omega+eV_L )\left[
N_{L}-R^e_{L}(\omega )+A^e_{L}(\omega )\right] & -h(\omega+eV_R )\left[ T_{LR}^e(\omega )-A_{LR}^e(\omega )\right]
\\- h(\omega+eV_L )\left[ T^e_{RL}(\omega )-A_{RL}^e(\omega )\right] & h(\omega+eV_R )\left[
N_{R}-R^e_{R}(\omega )+A^e_{R}(\omega )\right] \end{array}\right) ,
\label{diffcondmatrix2}
\end{equation}
where we note that $G_{\alpha\beta}$ is only a function of the voltage $V_\beta$.
The derivative of the Fermi-Dirac distribution function,
\begin{equation}
h(\omega )= -\frac{df(\omega)}{	d\omega } =\frac{1}{4k_{\mathrm{B}}T\cosh ^{2}(\omega/2k_{\mathrm{B}}T)},
\end{equation}
reduces to $h(\omega) = \delta(\omega)$ at zero temperature, and in that limit we find for the dimensionless conductance matrix
\begin{equation}
{\bf g}^0
= \left( \begin{array}{cc}
N_{L}-R^e_{L}(-eV_L)+A^e_{L}(-eV_L) & -[T_{LR}^e(-eV_R)-A_{LR}^e(-eV_R)]
\\- [T^e_{RL}(-eV_L)-A_{RL}^e(-eV_L)] &
N_{R}-R^e_{R}(-eV_R)+A^e_{R}(-eV_R)\end{array}\right).
\label{diffcondmatrix2zt}
\end{equation}

We then calculate the $S$-matrix using Eq.~(8) in the main text, using the toy Hamiltonian $H = E_0 \tau_z$.
Using the matrix ${\bf W}$ as given in Eq.~(9) in the main text (which assumes $N_L = N_R = 1$), we extract all transmission and reflection coefficients from ${\bf S}$ as written in (\ref{RATedef}), and find straightforwardly for the symmetric and antisymmetric components the following expressions:
\begin{align}
g^0_{LR,\mathrm{sym}}(\omega)
& = -4 \xi_L \xi_R \frac{E_0^2+\xi^2 - 8\, \mathrm{Re} [\xi^2_{LR}] + 2\gamma_L\gamma_R +\omega^2 }{\left( E_0^2+\xi^2-8\, \mathrm{Re} [\xi^2_{LR}] + 2\gamma_L\gamma_R -\omega^2\right)^2 + 4\gamma^2\omega^2}, \label{gLRsym} \\
g^0_{LR,\mathrm{asym}}(\omega)
& = {-}4 \xi_L \omega \frac{ 2 E_0 \gamma_R +8\,\mathrm{Im} [\xi^2_{LR}] }{ \left(E_0^2+\xi^2-8\,\mathrm{Re} [\xi^2_{LR}] + 2\gamma_L\gamma_R -\omega^2\right)^2+4\gamma^2\omega^2},
\label{gLRasym} \\
g^0_{LL,\mathrm{sym}}(\omega)
& = -\frac{\gamma_L\gamma_R - 4\,\mathrm{Re} [\xi^2_{LR}] }{\xi_L\xi_R} g^0_{LR,\mathrm{sym}}(\omega)
 +8\omega^2\frac{\gamma_L^2 - \xi_L^2 + 4\, \mathrm{Re} [\xi^2_{LR}]}{\left( E_0^2+\xi^2-8\,\mathrm{Re} [\xi^2_{LR}] + 2\gamma_L\gamma_R -\omega^2\right)^2 + 4\gamma^2\omega^2},
\label{gLLsym} \\
g^0_{LL,\mathrm{asym}}(\omega)
& = -g^0_{LR,\mathrm{asym}}(\omega),
\label{gLLasym}
\end{align}
where we introduced $\gamma_\alpha = \pi \nu |t_\alpha|^2 n_\alpha$, $\xi_\alpha = \pi \nu |t_\alpha|^2 q_\alpha$, $\gamma = \gamma_L+\gamma_R$, $\xi^2 = \xi^2_L + \xi^2_R$, and $\xi^2_{LR} = \pi^2\nu^2|t_Lt_R|^2 u_Lu_R^*v_L^*v_R$, with $n_\alpha = |u_\alpha|^2+|v_\alpha|^2$ and $q_\alpha = |u_\alpha|^2 - |v_\alpha|^2$.
%\end{widetext}
\twocolumngrid


\begin{thebibliography}{42}%
\makeatletter
\providecommand \@ifxundefined [1]{%
 \@ifx{#1\undefined}
}%
\providecommand \@ifnum [1]{%
 \ifnum #1\expandafter \@firstoftwo
 \else \expandafter \@secondoftwo
 \fi
}%
\providecommand \@ifx [1]{%
 \ifx #1\expandafter \@firstoftwo
 \else \expandafter \@secondoftwo
 \fi
}%
\providecommand \natexlab [1]{#1}%
\providecommand \enquote  [1]{``#1''}%
\providecommand \bibnamefont  [1]{#1}%
\providecommand \bibfnamefont [1]{#1}%
\providecommand \citenamefont [1]{#1}%
\providecommand \href@noop [0]{\@secondoftwo}%
\providecommand \href [0]{\begingroup \@sanitize@url \@href}%
\providecommand \@href[1]{\@@startlink{#1}\@@href}%
\providecommand \@@href[1]{\endgroup#1\@@endlink}%
\providecommand \@sanitize@url [0]{\catcode `\\12\catcode `\$12\catcode
  `\&12\catcode `\#12\catcode `\^12\catcode `\_12\catcode `\%12\relax}%
\providecommand \@@startlink[1]{}%
\providecommand \@@endlink[0]{}%
\providecommand \url  [0]{\begingroup\@sanitize@url \@url }%
\providecommand \@url [1]{\endgroup\@href {#1}{\urlprefix }}%
\providecommand \urlprefix  [0]{URL }%
\providecommand \Eprint [0]{\href }%
\providecommand \doibase [0]{http://dx.doi.org/}%
\providecommand \selectlanguage [0]{\@gobble}%
\providecommand \bibinfo  [0]{\@secondoftwo}%
\providecommand \bibfield  [0]{\@secondoftwo}%
\providecommand \translation [1]{[#1]}%
\providecommand \BibitemOpen [0]{}%
\providecommand \bibitemStop [0]{}%
\providecommand \bibitemNoStop [0]{.\EOS\space}%
\providecommand \EOS [0]{\spacefactor3000\relax}%
\providecommand \BibitemShut  [1]{\csname bibitem#1\endcsname}%
\let\auto@bib@innerbib\@empty
%</preamble>
\bibitem [{\citenamefont {Mourik}\ \emph {et~al.}(2012)\citenamefont {Mourik},
  \citenamefont {Zuo}, \citenamefont {Frolov}, \citenamefont {Plissard},
  \citenamefont {Bakkers},\ and\ \citenamefont {Kouwenhoven}}]{Mourik2012}%
  \BibitemOpen
  \bibfield  {author} {\bibinfo {author} {\bibfnamefont {V.}~\bibnamefont
  {Mourik}}, \bibinfo {author} {\bibfnamefont {K.}~\bibnamefont {Zuo}},
  \bibinfo {author} {\bibfnamefont {S.~M.}\ \bibnamefont {Frolov}}, \bibinfo
  {author} {\bibfnamefont {S.~R.}\ \bibnamefont {Plissard}}, \bibinfo {author}
  {\bibfnamefont {E.~P. A.~M.}\ \bibnamefont {Bakkers}}, \ and\ \bibinfo
  {author} {\bibfnamefont {L.~P.}\ \bibnamefont {Kouwenhoven}},\ }\href
  {\doibase 10.1126/science.1222360} {\bibfield  {journal} {\bibinfo  {journal}
  {Science}\ }\textbf {\bibinfo {volume} {336}},\ \bibinfo {pages} {1003}
  (\bibinfo {year} {2012})}\BibitemShut {NoStop}%
\bibitem [{\citenamefont {Das}\ \emph {et~al.}(2012)\citenamefont {Das},
  \citenamefont {Ronen}, \citenamefont {Most}, \citenamefont {Oreg},
  \citenamefont {Heiblum},\ and\ \citenamefont {Shtrikman}}]{Das2012}%
  \BibitemOpen
  \bibfield  {author} {\bibinfo {author} {\bibfnamefont {A.}~\bibnamefont
  {Das}}, \bibinfo {author} {\bibfnamefont {Y.}~\bibnamefont {Ronen}}, \bibinfo
  {author} {\bibfnamefont {Y.}~\bibnamefont {Most}}, \bibinfo {author}
  {\bibfnamefont {Y.}~\bibnamefont {Oreg}}, \bibinfo {author} {\bibfnamefont
  {M.}~\bibnamefont {Heiblum}}, \ and\ \bibinfo {author} {\bibfnamefont
  {H.}~\bibnamefont {Shtrikman}},\ }\href {\doibase 10.1038/nphys2479}
  {\bibfield  {journal} {\bibinfo  {journal} {Nat.~Phys.}\ }\textbf {\bibinfo
  {volume} {8}},\ \bibinfo {pages} {887} (\bibinfo {year} {2012})}\BibitemShut
  {NoStop}%
\bibitem [{\citenamefont {Deng}\ \emph {et~al.}(2012)\citenamefont {Deng},
  \citenamefont {Yu}, \citenamefont {Huang}, \citenamefont {Larsson},
  \citenamefont {Caroff},\ and\ \citenamefont {Xu}}]{Deng2012}%
  \BibitemOpen
  \bibfield  {author} {\bibinfo {author} {\bibfnamefont {M.~T.}\ \bibnamefont
  {Deng}}, \bibinfo {author} {\bibfnamefont {C.~L.}\ \bibnamefont {Yu}},
  \bibinfo {author} {\bibfnamefont {G.~Y.}\ \bibnamefont {Huang}}, \bibinfo
  {author} {\bibfnamefont {M.}~\bibnamefont {Larsson}}, \bibinfo {author}
  {\bibfnamefont {P.}~\bibnamefont {Caroff}}, \ and\ \bibinfo {author}
  {\bibfnamefont {H.~Q.}\ \bibnamefont {Xu}},\ }\href {\doibase
  10.1021/nl303758w} {\bibfield  {journal} {\bibinfo  {journal} {Nano Lett.}\
  }\textbf {\bibinfo {volume} {12}},\ \bibinfo {pages} {6414} (\bibinfo {year}
  {2012})}\BibitemShut {NoStop}%
\bibitem [{\citenamefont {Deng}\ \emph {et~al.}(2016)\citenamefont {Deng},
  \citenamefont {Vaitiek}, \citenamefont {Hansen}, \citenamefont {Danon},
  \citenamefont {Leijnse}, \citenamefont {Flensberg}, \citenamefont
  {Krogstrup},\ and\ \citenamefont {Marcus}}]{Deng2016}%
  \BibitemOpen
  \bibfield  {author} {\bibinfo {author} {\bibfnamefont {M.~T.}\ \bibnamefont
  {Deng}}, \bibinfo {author} {\bibfnamefont {S.}~\bibnamefont {Vaitiek}},
  \bibinfo {author} {\bibfnamefont {E.~B.}\ \bibnamefont {Hansen}}, \bibinfo
  {author} {\bibfnamefont {J.}~\bibnamefont {Danon}}, \bibinfo {author}
  {\bibfnamefont {M.}~\bibnamefont {Leijnse}}, \bibinfo {author} {\bibfnamefont
  {K.}~\bibnamefont {Flensberg}}, \bibinfo {author} {\bibfnamefont
  {P.}~\bibnamefont {Krogstrup}}, \ and\ \bibinfo {author} {\bibfnamefont
  {C.~M.}\ \bibnamefont {Marcus}},\ }\href {\doibase 10.1126/science.aaf3961}
  {\bibfield  {journal} {\bibinfo  {journal} {Science}\ }\textbf {\bibinfo
  {volume} {354}},\ \bibinfo {pages} {1557} (\bibinfo {year}
  {2016})}\BibitemShut {NoStop}%
\bibitem [{\citenamefont {Nichele}\ \emph {et~al.}(2017)\citenamefont
  {Nichele}, \citenamefont {Drachmann}, \citenamefont {Whiticar}, \citenamefont
  {O'Farrell}, \citenamefont {Suominen}, \citenamefont {Fornieri},
  \citenamefont {Wang}, \citenamefont {Gardner}, \citenamefont {Thomas},
  \citenamefont {Hatke}, \citenamefont {Krogstrup}, \citenamefont {Manfra},
  \citenamefont {Flensberg},\ and\ \citenamefont {Marcus}}]{Nichele2017}%
  \BibitemOpen
  \bibfield  {author} {\bibinfo {author} {\bibfnamefont {F.}~\bibnamefont
  {Nichele}}, \bibinfo {author} {\bibfnamefont {A.~C.~C.}\ \bibnamefont
  {Drachmann}}, \bibinfo {author} {\bibfnamefont {A.~M.}\ \bibnamefont
  {Whiticar}}, \bibinfo {author} {\bibfnamefont {E.~C.~T.}\ \bibnamefont
  {O'Farrell}}, \bibinfo {author} {\bibfnamefont {H.~J.}\ \bibnamefont
  {Suominen}}, \bibinfo {author} {\bibfnamefont {A.}~\bibnamefont {Fornieri}},
  \bibinfo {author} {\bibfnamefont {T.}~\bibnamefont {Wang}}, \bibinfo {author}
  {\bibfnamefont {G.~C.}\ \bibnamefont {Gardner}}, \bibinfo {author}
  {\bibfnamefont {C.}~\bibnamefont {Thomas}}, \bibinfo {author} {\bibfnamefont
  {A.~T.}\ \bibnamefont {Hatke}}, \bibinfo {author} {\bibfnamefont
  {P.}~\bibnamefont {Krogstrup}}, \bibinfo {author} {\bibfnamefont {M.~J.}\
  \bibnamefont {Manfra}}, \bibinfo {author} {\bibfnamefont {K.}~\bibnamefont
  {Flensberg}}, \ and\ \bibinfo {author} {\bibfnamefont {C.~M.}\ \bibnamefont
  {Marcus}},\ }\href {\doibase 10.1103/PhysRevLett.119.136803} {\bibfield
  {journal} {\bibinfo  {journal} {Phys.~Rev.~Lett.}\ }\textbf {\bibinfo
  {volume} {119}},\ \bibinfo {pages} {136803} (\bibinfo {year}
  {2017})}\BibitemShut {NoStop}%
\bibitem [{\citenamefont {Zhang}\ \emph {et~al.}(2018)\citenamefont {Zhang},
  \citenamefont {Liu}, \citenamefont {Gazibegovic}, \citenamefont {Xu},
  \citenamefont {Logan}, \citenamefont {Wang}, \citenamefont {van Loo},
  \citenamefont {Bommer}, \citenamefont {de~Moor}, \citenamefont {Car},
  \citenamefont {het Veld}, \citenamefont {van Veldhoven}, \citenamefont
  {Koelling}, \citenamefont {Verheijen}, \citenamefont {Pendharkar},
  \citenamefont {Pennachio}, \citenamefont {Shojaei}, \citenamefont {Lee},
  \citenamefont {Palmstrom}, \citenamefont {Bakkers}, \citenamefont {Sarma},\
  and\ \citenamefont {Kouwenhoven}}]{Zhang2018}%
  \BibitemOpen
  \bibfield  {author} {\bibinfo {author} {\bibfnamefont {H.}~\bibnamefont
  {Zhang}}, \bibinfo {author} {\bibfnamefont {C.-X.}\ \bibnamefont {Liu}},
  \bibinfo {author} {\bibfnamefont {S.}~\bibnamefont {Gazibegovic}}, \bibinfo
  {author} {\bibfnamefont {D.}~\bibnamefont {Xu}}, \bibinfo {author}
  {\bibfnamefont {J.~A.}\ \bibnamefont {Logan}}, \bibinfo {author}
  {\bibfnamefont {G.}~\bibnamefont {Wang}}, \bibinfo {author} {\bibfnamefont
  {N.}~\bibnamefont {van Loo}}, \bibinfo {author} {\bibfnamefont {J.~D.~S.}\
  \bibnamefont {Bommer}}, \bibinfo {author} {\bibfnamefont {M.~W.~A.}\
  \bibnamefont {de~Moor}}, \bibinfo {author} {\bibfnamefont {D.}~\bibnamefont
  {Car}}, \bibinfo {author} {\bibfnamefont {R.~L. M.~O.}\ \bibnamefont {het
  Veld}}, \bibinfo {author} {\bibfnamefont {P.~J.}\ \bibnamefont {van
  Veldhoven}}, \bibinfo {author} {\bibfnamefont {S.}~\bibnamefont {Koelling}},
  \bibinfo {author} {\bibfnamefont {M.~A.}\ \bibnamefont {Verheijen}}, \bibinfo
  {author} {\bibfnamefont {M.}~\bibnamefont {Pendharkar}}, \bibinfo {author}
  {\bibfnamefont {D.~J.}\ \bibnamefont {Pennachio}}, \bibinfo {author}
  {\bibfnamefont {B.}~\bibnamefont {Shojaei}}, \bibinfo {author} {\bibfnamefont
  {J.~S.}\ \bibnamefont {Lee}}, \bibinfo {author} {\bibfnamefont {C.~J.}\
  \bibnamefont {Palmstrom}}, \bibinfo {author} {\bibfnamefont {E.~P. A.~M.}\
  \bibnamefont {Bakkers}}, \bibinfo {author} {\bibfnamefont {S.~D.}\
  \bibnamefont {Sarma}}, \ and\ \bibinfo {author} {\bibfnamefont {L.~P.}\
  \bibnamefont {Kouwenhoven}},\ }\href {\doibase 10.1038/nature26142}
  {\bibfield  {journal} {\bibinfo  {journal} {Nature}\ }\textbf {\bibinfo
  {volume} {556}},\ \bibinfo {pages} {74} (\bibinfo {year} {2018})}\BibitemShut
  {NoStop}%
\bibitem [{Vai()}]{Vaitiekenas2018b}%
  \BibitemOpen
  \href@noop {} {}\bibinfo {note} {S. Vaitiekenas, M.-T. Deng, P. Krogstrup, C.
  M. Marcus, \textit{Flux-induced Majorana modes in full-shell nanowires},
  \href{https://arxiv.org/abs/1809.05513}{ arXiv:1809.05513}}\BibitemShut
  {NoStop}%
\bibitem [{\citenamefont {Sengupta}\ \emph {et~al.}(2001)\citenamefont
  {Sengupta}, \citenamefont {\v{Z}uti\'{c}}, \citenamefont {Kwon},
  \citenamefont {Yakovenko},\ and\ \citenamefont {{Das Sarma}}}]{Sengupta2001}%
  \BibitemOpen
  \bibfield  {author} {\bibinfo {author} {\bibfnamefont {K.}~\bibnamefont
  {Sengupta}}, \bibinfo {author} {\bibfnamefont {I.}~\bibnamefont
  {\v{Z}uti\'{c}}}, \bibinfo {author} {\bibfnamefont {H.-J.}\ \bibnamefont
  {Kwon}}, \bibinfo {author} {\bibfnamefont {V.~M.}\ \bibnamefont {Yakovenko}},
  \ and\ \bibinfo {author} {\bibfnamefont {S.}~\bibnamefont {{Das Sarma}}},\
  }\href {\doibase 10.1103/PhysRevB.63.144531} {\bibfield  {journal} {\bibinfo
  {journal} {Phys.~Rev.~B}\ }\textbf {\bibinfo {volume} {63}},\ \bibinfo
  {pages} {144531} (\bibinfo {year} {2001})}\BibitemShut {NoStop}%
\bibitem [{\citenamefont {Law}\ \emph {et~al.}(2009)\citenamefont {Law},
  \citenamefont {Lee},\ and\ \citenamefont {Ng}}]{Law2009}%
  \BibitemOpen
  \bibfield  {author} {\bibinfo {author} {\bibfnamefont {K.~T.}\ \bibnamefont
  {Law}}, \bibinfo {author} {\bibfnamefont {P.~A.}\ \bibnamefont {Lee}}, \ and\
  \bibinfo {author} {\bibfnamefont {T.~K.}\ \bibnamefont {Ng}},\ }\href
  {\doibase 10.1103/PhysRevLett.103.237001} {\bibfield  {journal} {\bibinfo
  {journal} {Phys.~Rev.~Lett.}\ }\textbf {\bibinfo {volume} {103}},\ \bibinfo
  {pages} {237001} (\bibinfo {year} {2009})}\BibitemShut {NoStop}%
\bibitem [{\citenamefont {Flensberg}(2010)}]{Flensberg2010}%
  \BibitemOpen
  \bibfield  {author} {\bibinfo {author} {\bibfnamefont {K.}~\bibnamefont
  {Flensberg}},\ }\href {\doibase 10.1103/PhysRevB.82.180516} {\bibfield
  {journal} {\bibinfo  {journal} {Phys.~Rev.~B}\ }\textbf {\bibinfo {volume}
  {82}},\ \bibinfo {pages} {180516(R)} (\bibinfo {year} {2010})}\BibitemShut
  {NoStop}%
\bibitem [{\citenamefont {Hansen}\ \emph {et~al.}(2016)\citenamefont {Hansen},
  \citenamefont {Danon},\ and\ \citenamefont {Flensberg}}]{Hansen2016}%
  \BibitemOpen
  \bibfield  {author} {\bibinfo {author} {\bibfnamefont {E.~B.}\ \bibnamefont
  {Hansen}}, \bibinfo {author} {\bibfnamefont {J.}~\bibnamefont {Danon}}, \
  and\ \bibinfo {author} {\bibfnamefont {K.}~\bibnamefont {Flensberg}},\ }\href
  {\doibase 10.1103/PhysRevB.93.094501} {\bibfield  {journal} {\bibinfo
  {journal} {Phys.~Rev.~B}\ }\textbf {\bibinfo {volume} {93}},\ \bibinfo
  {pages} {094501(R)} (\bibinfo {year} {2016})}\BibitemShut {NoStop}%
\bibitem [{\citenamefont {Leijnse}\ and\ \citenamefont
  {Flensberg}(2011)}]{Leijnse2011}%
  \BibitemOpen
  \bibfield  {author} {\bibinfo {author} {\bibfnamefont {M.}~\bibnamefont
  {Leijnse}}\ and\ \bibinfo {author} {\bibfnamefont {K.}~\bibnamefont
  {Flensberg}},\ }\href {\doibase 10.1103/PhysRevB.84.140501} {\bibfield
  {journal} {\bibinfo  {journal} {Phys.~Rev.~B}\ }\textbf {\bibinfo {volume}
  {84}},\ \bibinfo {pages} {140501(R)} (\bibinfo {year} {2011})}\BibitemShut
  {NoStop}%
\bibitem [{\citenamefont {Prada}\ \emph {et~al.}(2017)\citenamefont {Prada},
  \citenamefont {Aguado},\ and\ \citenamefont {San-Jose}}]{Prada2017}%
  \BibitemOpen
  \bibfield  {author} {\bibinfo {author} {\bibfnamefont {E.}~\bibnamefont
  {Prada}}, \bibinfo {author} {\bibfnamefont {R.}~\bibnamefont {Aguado}}, \
  and\ \bibinfo {author} {\bibfnamefont {P.}~\bibnamefont {San-Jose}},\ }\href
  {\doibase 10.1103/PhysRevB.96.085418} {\bibfield  {journal} {\bibinfo
  {journal} {Phys.~Rev.~B}\ }\textbf {\bibinfo {volume} {96}},\ \bibinfo
  {pages} {085418} (\bibinfo {year} {2017})}\BibitemShut {NoStop}%
\bibitem [{\citenamefont {Clarke}(2017)}]{Clarke2017}%
  \BibitemOpen
  \bibfield  {author} {\bibinfo {author} {\bibfnamefont {D.~J.}\ \bibnamefont
  {Clarke}},\ }\href {\doibase 10.1103/PhysRevB.96.201109} {\bibfield
  {journal} {\bibinfo  {journal} {Phys.~Rev.~B}\ }\textbf {\bibinfo {volume}
  {96}},\ \bibinfo {pages} {201109(R)} (\bibinfo {year} {2017})}\BibitemShut
  {NoStop}%
\bibitem [{\citenamefont {Deng}\ \emph {et~al.}(2018)\citenamefont {Deng},
  \citenamefont {Vaitiekenas}, \citenamefont {Prada}, \citenamefont {San-Jose},
  \citenamefont {Nyg{\aa}rd}, \citenamefont {Krogstrup}, \citenamefont
  {Aguado},\ and\ \citenamefont {Marcus}}]{Deng2018}%
  \BibitemOpen
  \bibfield  {author} {\bibinfo {author} {\bibfnamefont {M.~T.}\ \bibnamefont
  {Deng}}, \bibinfo {author} {\bibfnamefont {S.}~\bibnamefont {Vaitiekenas}},
  \bibinfo {author} {\bibfnamefont {E.}~\bibnamefont {Prada}}, \bibinfo
  {author} {\bibfnamefont {P.}~\bibnamefont {San-Jose}}, \bibinfo {author}
  {\bibfnamefont {J.}~\bibnamefont {Nyg{\aa}rd}}, \bibinfo {author}
  {\bibfnamefont {P.}~\bibnamefont {Krogstrup}}, \bibinfo {author}
  {\bibfnamefont {R.}~\bibnamefont {Aguado}}, \ and\ \bibinfo {author}
  {\bibfnamefont {C.~M.}\ \bibnamefont {Marcus}},\ }\href {\doibase
  10.1103/PhysRevB.98.085125} {\bibfield  {journal} {\bibinfo  {journal}
  {Phys.~Rev.~B}\ }\textbf {\bibinfo {volume} {98}},\ \bibinfo {pages} {085125}
  (\bibinfo {year} {2018})}\BibitemShut {NoStop}%
\bibitem [{\citenamefont {Kells}\ \emph {et~al.}(2012)\citenamefont {Kells},
  \citenamefont {Meidan},\ and\ \citenamefont {Brouwer}}]{Kell2012}%
  \BibitemOpen
  \bibfield  {author} {\bibinfo {author} {\bibfnamefont {G.}~\bibnamefont
  {Kells}}, \bibinfo {author} {\bibfnamefont {D.}~\bibnamefont {Meidan}}, \
  and\ \bibinfo {author} {\bibfnamefont {P.~W.}\ \bibnamefont {Brouwer}},\
  }\href {\doibase 10.1103/PhysRevB.86.100503} {\bibfield  {journal} {\bibinfo
  {journal} {Phys.~Rev.~B}\ }\textbf {\bibinfo {volume} {86}},\ \bibinfo
  {pages} {100503(R)} (\bibinfo {year} {2012})}\BibitemShut {NoStop}%
\bibitem [{\citenamefont {Prada}\ \emph {et~al.}(2012)\citenamefont {Prada},
  \citenamefont {San-Jose},\ and\ \citenamefont {Aguado}}]{Prada2012}%
  \BibitemOpen
  \bibfield  {author} {\bibinfo {author} {\bibfnamefont {E.}~\bibnamefont
  {Prada}}, \bibinfo {author} {\bibfnamefont {P.}~\bibnamefont {San-Jose}}, \
  and\ \bibinfo {author} {\bibfnamefont {R.}~\bibnamefont {Aguado}},\ }\href
  {\doibase 10.1103/PhysRevB.86.180503} {\bibfield  {journal} {\bibinfo
  {journal} {Phys.~Rev.~B}\ }\textbf {\bibinfo {volume} {86}},\ \bibinfo
  {pages} {180503(R)} (\bibinfo {year} {2012})}\BibitemShut {NoStop}%
\bibitem [{Fle(2018)}]{Fleckenstein2018}%
  \BibitemOpen
  \href {\doibase 10.1103/PhysRevB.97.155425} {\bibfield  {journal} {\bibinfo
  {journal} {Phys.~Rev.~B}\ }\textbf {\bibinfo {volume} {97}},\ \bibinfo
  {pages} {155425} (\bibinfo {year} {2018})}\BibitemShut {NoStop}%
\bibitem [{\citenamefont {Liu}\ \emph {et~al.}(2017)\citenamefont {Liu},
  \citenamefont {Sau}, \citenamefont {Stanescu},\ and\ \citenamefont {{Das
  Sarma}}}]{Liu2017}%
  \BibitemOpen
  \bibfield  {author} {\bibinfo {author} {\bibfnamefont {C.~X.}\ \bibnamefont
  {Liu}}, \bibinfo {author} {\bibfnamefont {J.~D.}\ \bibnamefont {Sau}},
  \bibinfo {author} {\bibfnamefont {T.~D.}\ \bibnamefont {Stanescu}}, \ and\
  \bibinfo {author} {\bibfnamefont {S.}~\bibnamefont {{Das Sarma}}},\ }\href
  {\doibase 10.1103/PhysRevB.96.075161} {\bibfield  {journal} {\bibinfo
  {journal} {Phys.~Rev.~B}\ }\textbf {\bibinfo {volume} {96}},\ \bibinfo
  {pages} {075161} (\bibinfo {year} {2017})}\BibitemShut {NoStop}%
\bibitem [{Vui()}]{Vuik2018}%
  \BibitemOpen
  \href@noop {} {}\bibinfo {note} {A. Vuik, B. Nijholt, A. R. Akhmerov, M.
  Wimmer,
  \href{https://arxiv.org/abs/1806.02801}{arXiv:1806:02801}}\BibitemShut
  {NoStop}%
\bibitem [{\citenamefont {Albrecht}\ \emph {et~al.}(2016)\citenamefont
  {Albrecht}, \citenamefont {Higginbotham}, \citenamefont {Madsen},
  \citenamefont {Kuemmeth}, \citenamefont {Jespersen}, \citenamefont
  {Nyg{\aa}rd}, \citenamefont {Krogstrup},\ and\ \citenamefont
  {Marcus}}]{Albrecht2016}%
  \BibitemOpen
  \bibfield  {author} {\bibinfo {author} {\bibfnamefont {S.~M.}\ \bibnamefont
  {Albrecht}}, \bibinfo {author} {\bibfnamefont {A.~P.}\ \bibnamefont
  {Higginbotham}}, \bibinfo {author} {\bibfnamefont {M.}~\bibnamefont
  {Madsen}}, \bibinfo {author} {\bibfnamefont {F.}~\bibnamefont {Kuemmeth}},
  \bibinfo {author} {\bibfnamefont {T.~S.}\ \bibnamefont {Jespersen}}, \bibinfo
  {author} {\bibfnamefont {J.}~\bibnamefont {Nyg{\aa}rd}}, \bibinfo {author}
  {\bibfnamefont {P.}~\bibnamefont {Krogstrup}}, \ and\ \bibinfo {author}
  {\bibfnamefont {C.~M.}\ \bibnamefont {Marcus}},\ }\href {\doibase
  10.1038/nature17162} {\bibfield  {journal} {\bibinfo  {journal} {Nature}\
  }\textbf {\bibinfo {volume} {531}},\ \bibinfo {pages} {206} (\bibinfo {year}
  {2016})}\BibitemShut {NoStop}%
\bibitem [{\citenamefont {Albrecht}\ \emph {et~al.}(2017)\citenamefont
  {Albrecht}, \citenamefont {Hansen}, \citenamefont {Higginbotham},
  \citenamefont {Kuemmeth}, \citenamefont {Jespersen}, \citenamefont
  {Nyg{\aa}rd}, \citenamefont {Krogstrup}, \citenamefont {Danon}, \citenamefont
  {Flensberg},\ and\ \citenamefont {Marcus}}]{Albrecht2017}%
  \BibitemOpen
  \bibfield  {author} {\bibinfo {author} {\bibfnamefont {S.~M.}\ \bibnamefont
  {Albrecht}}, \bibinfo {author} {\bibfnamefont {E.~B.}\ \bibnamefont
  {Hansen}}, \bibinfo {author} {\bibfnamefont {A.~P.}\ \bibnamefont
  {Higginbotham}}, \bibinfo {author} {\bibfnamefont {F.}~\bibnamefont
  {Kuemmeth}}, \bibinfo {author} {\bibfnamefont {T.~S.}\ \bibnamefont
  {Jespersen}}, \bibinfo {author} {\bibfnamefont {J.}~\bibnamefont
  {Nyg{\aa}rd}}, \bibinfo {author} {\bibfnamefont {P.}~\bibnamefont
  {Krogstrup}}, \bibinfo {author} {\bibfnamefont {J.}~\bibnamefont {Danon}},
  \bibinfo {author} {\bibfnamefont {K.}~\bibnamefont {Flensberg}}, \ and\
  \bibinfo {author} {\bibfnamefont {C.~M.}\ \bibnamefont {Marcus}},\ }\href
  {\doibase 10.1103/PhysRevLett.118.137701} {\bibfield  {journal} {\bibinfo
  {journal} {Phys.~Rev.~Lett.}\ }\textbf {\bibinfo {volume} {118}},\ \bibinfo
  {pages} {137701} (\bibinfo {year} {2017})}\BibitemShut {NoStop}%
\bibitem [{\citenamefont {Vaitiekenas}\ \emph
  {et~al.}(2018{\natexlab{a}})\citenamefont {Vaitiekenas}, \citenamefont
  {Deng}, \citenamefont {Nyg{\aa}rd}, \citenamefont {Krogstrup},\ and\
  \citenamefont {Marcus}}]{Vaitiekenas2018}%
  \BibitemOpen
  \bibfield  {author} {\bibinfo {author} {\bibfnamefont {S.}~\bibnamefont
  {Vaitiekenas}}, \bibinfo {author} {\bibfnamefont {M.~T.}\ \bibnamefont
  {Deng}}, \bibinfo {author} {\bibfnamefont {J.}~\bibnamefont {Nyg{\aa}rd}},
  \bibinfo {author} {\bibfnamefont {P.}~\bibnamefont {Krogstrup}}, \ and\
  \bibinfo {author} {\bibfnamefont {C.~M.}\ \bibnamefont {Marcus}},\ }\href
  {\doibase 10.1103/PhysRevLett.121.037703} {\bibfield  {journal} {\bibinfo
  {journal} {Phys.~Rev.~Lett.}\ }\textbf {\bibinfo {volume} {121}},\ \bibinfo
  {pages} {037703} (\bibinfo {year} {2018}{\natexlab{a}})}\BibitemShut
  {NoStop}%
\bibitem [{\citenamefont {Vaitiekenas}\ \emph
  {et~al.}(2018{\natexlab{b}})\citenamefont {Vaitiekenas}, \citenamefont
  {Whiticar}, \citenamefont {Deng}, \citenamefont {Krizek}, \citenamefont
  {Sestoft}, \citenamefont {Palmstr{\o}m}, \citenamefont {Marti-Sanchez},
  \citenamefont {Arbiol}, \citenamefont {Krogstrup}, \citenamefont {Casparis},\
  and\ \citenamefont {Marcus}}]{Vaitiekenas2018c}%
  \BibitemOpen
  \bibfield  {author} {\bibinfo {author} {\bibfnamefont {S.}~\bibnamefont
  {Vaitiekenas}}, \bibinfo {author} {\bibfnamefont {A.~M.}\ \bibnamefont
  {Whiticar}}, \bibinfo {author} {\bibfnamefont {M.~T.}\ \bibnamefont {Deng}},
  \bibinfo {author} {\bibfnamefont {F.}~\bibnamefont {Krizek}}, \bibinfo
  {author} {\bibfnamefont {J.~E.}\ \bibnamefont {Sestoft}}, \bibinfo {author}
  {\bibfnamefont {C.~J.}\ \bibnamefont {Palmstr{\o}m}}, \bibinfo {author}
  {\bibfnamefont {S.}~\bibnamefont {Marti-Sanchez}}, \bibinfo {author}
  {\bibfnamefont {J.}~\bibnamefont {Arbiol}}, \bibinfo {author} {\bibfnamefont
  {P.}~\bibnamefont {Krogstrup}}, \bibinfo {author} {\bibfnamefont
  {L.}~\bibnamefont {Casparis}}, \ and\ \bibinfo {author} {\bibfnamefont
  {C.~M.}\ \bibnamefont {Marcus}},\ }\href {\doibase
  10.1103/PhysRevLett.121.147701} {\bibfield  {journal} {\bibinfo  {journal}
  {Phys.~Rev.~Lett.}\ }\textbf {\bibinfo {volume} {121}},\ \bibinfo {pages}
  {147701} (\bibinfo {year} {2018}{\natexlab{b}})}\BibitemShut {NoStop}%
\bibitem [{\citenamefont {Shen}\ \emph {et~al.}(2018)\citenamefont {Shen},
  \citenamefont {Heedt}, \citenamefont {Borsoi}, \citenamefont {van Heck},
  \citenamefont {Gazibegovic}, \citenamefont {{Op Het Veld}}, \citenamefont
  {Car}, \citenamefont {Logan}, \citenamefont {Pendharkar}, \citenamefont
  {Ramakers}, \citenamefont {Wang}, \citenamefont {Xu}, \citenamefont {Bouman},
  \citenamefont {Geresdi}, \citenamefont {Palmstr{\o}m}, \citenamefont
  {Bakkers},\ and\ \citenamefont {Kouwenhoven}}]{Shen2018}%
  \BibitemOpen
  \bibfield  {author} {\bibinfo {author} {\bibfnamefont {J.}~\bibnamefont
  {Shen}}, \bibinfo {author} {\bibfnamefont {S.}~\bibnamefont {Heedt}},
  \bibinfo {author} {\bibfnamefont {F.}~\bibnamefont {Borsoi}}, \bibinfo
  {author} {\bibfnamefont {B.}~\bibnamefont {van Heck}}, \bibinfo {author}
  {\bibfnamefont {S.}~\bibnamefont {Gazibegovic}}, \bibinfo {author}
  {\bibfnamefont {R.~L.}\ \bibnamefont {{Op Het Veld}}}, \bibinfo {author}
  {\bibfnamefont {D.}~\bibnamefont {Car}}, \bibinfo {author} {\bibfnamefont
  {J.~A.}\ \bibnamefont {Logan}}, \bibinfo {author} {\bibfnamefont
  {M.}~\bibnamefont {Pendharkar}}, \bibinfo {author} {\bibfnamefont {S.~J.}\
  \bibnamefont {Ramakers}}, \bibinfo {author} {\bibfnamefont {G.}~\bibnamefont
  {Wang}}, \bibinfo {author} {\bibfnamefont {D.}~\bibnamefont {Xu}}, \bibinfo
  {author} {\bibfnamefont {D.}~\bibnamefont {Bouman}}, \bibinfo {author}
  {\bibfnamefont {A.}~\bibnamefont {Geresdi}}, \bibinfo {author} {\bibfnamefont
  {C.~J.}\ \bibnamefont {Palmstr{\o}m}}, \bibinfo {author} {\bibfnamefont
  {E.~P.}\ \bibnamefont {Bakkers}}, \ and\ \bibinfo {author} {\bibfnamefont
  {L.~P.}\ \bibnamefont {Kouwenhoven}},\ }\href {\doibase
  10.1038/s41467-018-07279-7} {\bibfield  {journal} {\bibinfo  {journal}
  {Nature Communications}\ }\textbf {\bibinfo {volume} {9}},\ \bibinfo {pages}
  {4801} (\bibinfo {year} {2018})}\BibitemShut {NoStop}%
\bibitem [{\citenamefont {{van Heck}}\ \emph {et~al.}(2016)\citenamefont {{van
  Heck}}, \citenamefont {Lutchyn},\ and\ \citenamefont
  {Glazman}}]{VanHeck2016}%
  \BibitemOpen
  \bibfield  {author} {\bibinfo {author} {\bibfnamefont {B.}~\bibnamefont {{van
  Heck}}}, \bibinfo {author} {\bibfnamefont {R.~M.}\ \bibnamefont {Lutchyn}}, \
  and\ \bibinfo {author} {\bibfnamefont {L.~I.}\ \bibnamefont {Glazman}},\
  }\href {\doibase 10.1103/PhysRevB.93.235431} {\bibfield  {journal} {\bibinfo
  {journal} {Phys.~Rev.~B}\ }\textbf {\bibinfo {volume} {93}},\ \bibinfo
  {pages} {235431} (\bibinfo {year} {2016})}\BibitemShut {NoStop}%
\bibitem [{\citenamefont {Chiu}\ \emph {et~al.}(2017)\citenamefont {Chiu},
  \citenamefont {Sau},\ and\ \citenamefont {{Das Sarma}}}]{Chiu2017}%
  \BibitemOpen
  \bibfield  {author} {\bibinfo {author} {\bibfnamefont {C.~K.}\ \bibnamefont
  {Chiu}}, \bibinfo {author} {\bibfnamefont {J.~D.}\ \bibnamefont {Sau}}, \
  and\ \bibinfo {author} {\bibfnamefont {S.}~\bibnamefont {{Das Sarma}}},\
  }\href {\doibase 10.1103/PhysRevB.96.054504} {\bibfield  {journal} {\bibinfo
  {journal} {Phys.~Rev.~B}\ }\textbf {\bibinfo {volume} {96}},\ \bibinfo
  {pages} {054504} (\bibinfo {year} {2017})}\BibitemShut {NoStop}%
\bibitem [{\citenamefont {Higginbotham}\ \emph {et~al.}(2015)\citenamefont
  {Higginbotham}, \citenamefont {Albrecht}, \citenamefont {Kir{\v{s}}anskas},
  \citenamefont {Chang}, \citenamefont {Kuemmeth}, \citenamefont {Krogstrup},
  \citenamefont {Jespersen}, \citenamefont {Nyg{\aa}rd}, \citenamefont
  {Flensberg},\ and\ \citenamefont {Marcus}}]{Higginbotham2015}%
  \BibitemOpen
  \bibfield  {author} {\bibinfo {author} {\bibfnamefont {A.~P.}\ \bibnamefont
  {Higginbotham}}, \bibinfo {author} {\bibfnamefont {S.}~\bibnamefont
  {Albrecht}}, \bibinfo {author} {\bibfnamefont {G.}~\bibnamefont
  {Kir{\v{s}}anskas}}, \bibinfo {author} {\bibfnamefont {W.}~\bibnamefont
  {Chang}}, \bibinfo {author} {\bibfnamefont {F.}~\bibnamefont {Kuemmeth}},
  \bibinfo {author} {\bibfnamefont {P.}~\bibnamefont {Krogstrup}}, \bibinfo
  {author} {\bibfnamefont {T.}~\bibnamefont {Jespersen}}, \bibinfo {author}
  {\bibfnamefont {J.}~\bibnamefont {Nyg{\aa}rd}}, \bibinfo {author}
  {\bibfnamefont {K.}~\bibnamefont {Flensberg}}, \ and\ \bibinfo {author}
  {\bibfnamefont {C.}~\bibnamefont {Marcus}},\ }\href {\doibase
  10.1038/nphys3461} {\bibfield  {journal} {\bibinfo  {journal} {Nat.~Phys.}\
  }\textbf {\bibinfo {volume} {11}},\ \bibinfo {pages} {107} (\bibinfo {year}
  {2015})}\BibitemShut {NoStop}%
\bibitem [{\citenamefont {Gramich}\ \emph {et~al.}(2017)\citenamefont
  {Gramich}, \citenamefont {Baumgartner},\ and\ \citenamefont
  {Sch{\"{o}}nenberger}}]{Gramich2017}%
  \BibitemOpen
  \bibfield  {author} {\bibinfo {author} {\bibfnamefont {J.}~\bibnamefont
  {Gramich}}, \bibinfo {author} {\bibfnamefont {A.}~\bibnamefont
  {Baumgartner}}, \ and\ \bibinfo {author} {\bibfnamefont {C.}~\bibnamefont
  {Sch{\"{o}}nenberger}},\ }\href {\doibase 10.1103/PhysRevB.96.195418}
  {\bibfield  {journal} {\bibinfo  {journal} {Phys.~Rev.~B}\ }\textbf {\bibinfo
  {volume} {96}},\ \bibinfo {pages} {195418} (\bibinfo {year}
  {2017})}\BibitemShut {NoStop}%
\bibitem [{\citenamefont {Hofstetter}\ \emph {et~al.}(2009)\citenamefont
  {Hofstetter}, \citenamefont {Csonka}, \citenamefont {Nyg{\aa}rd},\ and\
  \citenamefont {Sch{\"{o}}nenberger}}]{Hofstetter2009}%
  \BibitemOpen
  \bibfield  {author} {\bibinfo {author} {\bibfnamefont {L.}~\bibnamefont
  {Hofstetter}}, \bibinfo {author} {\bibfnamefont {S.}~\bibnamefont {Csonka}},
  \bibinfo {author} {\bibfnamefont {J.}~\bibnamefont {Nyg{\aa}rd}}, \ and\
  \bibinfo {author} {\bibfnamefont {C.}~\bibnamefont {Sch{\"{o}}nenberger}},\
  }\href {\doibase 10.1038/nature08432} {\bibfield  {journal} {\bibinfo
  {journal} {Nature}\ }\textbf {\bibinfo {volume} {461}},\ \bibinfo {pages}
  {960} (\bibinfo {year} {2009})}\BibitemShut {NoStop}%
\bibitem [{\citenamefont {Herrmann}\ \emph {et~al.}(2010)\citenamefont
  {Herrmann}, \citenamefont {Portier}, \citenamefont {Roche}, \citenamefont
  {Yeyati}, \citenamefont {Kontos},\ and\ \citenamefont
  {Strunk}}]{Herrmann2010}%
  \BibitemOpen
  \bibfield  {author} {\bibinfo {author} {\bibfnamefont {L.~G.}\ \bibnamefont
  {Herrmann}}, \bibinfo {author} {\bibfnamefont {F.}~\bibnamefont {Portier}},
  \bibinfo {author} {\bibfnamefont {P.}~\bibnamefont {Roche}}, \bibinfo
  {author} {\bibfnamefont {A.~L.}\ \bibnamefont {Yeyati}}, \bibinfo {author}
  {\bibfnamefont {T.}~\bibnamefont {Kontos}}, \ and\ \bibinfo {author}
  {\bibfnamefont {C.}~\bibnamefont {Strunk}},\ }\href {\doibase
  10.1103/PhysRevLett.104.026801} {\bibfield  {journal} {\bibinfo  {journal}
  {Phys.~Rev.~Lett.}\ }\textbf {\bibinfo {volume} {104}},\ \bibinfo {pages}
  {026801} (\bibinfo {year} {2010})}\BibitemShut {NoStop}%
\bibitem [{\citenamefont {Schindele}\ \emph {et~al.}(2012)\citenamefont
  {Schindele}, \citenamefont {Baumgartner},\ and\ \citenamefont
  {Sch{\"{o}}nenberger}}]{Schindele2012}%
  \BibitemOpen
  \bibfield  {author} {\bibinfo {author} {\bibfnamefont {J.}~\bibnamefont
  {Schindele}}, \bibinfo {author} {\bibfnamefont {A.}~\bibnamefont
  {Baumgartner}}, \ and\ \bibinfo {author} {\bibfnamefont {C.}~\bibnamefont
  {Sch{\"{o}}nenberger}},\ }\href {\doibase 10.1103/PhysRevLett.109.157002}
  {\bibfield  {journal} {\bibinfo  {journal} {Phys.~Rev.~Lett.}\ }\textbf
  {\bibinfo {volume} {109}},\ \bibinfo {pages} {157002} (\bibinfo {year}
  {2012})}\BibitemShut {NoStop}%
\bibitem [{\citenamefont {Recher}\ \emph {et~al.}(2001)\citenamefont {Recher},
  \citenamefont {Sukhorukov},\ and\ \citenamefont {Loss}}]{Recher2001}%
  \BibitemOpen
  \bibfield  {author} {\bibinfo {author} {\bibfnamefont {P.}~\bibnamefont
  {Recher}}, \bibinfo {author} {\bibfnamefont {E.~V.}\ \bibnamefont
  {Sukhorukov}}, \ and\ \bibinfo {author} {\bibfnamefont {D.}~\bibnamefont
  {Loss}},\ }\href {\doibase 10.1103/PhysRevB.63.165314} {\bibfield  {journal}
  {\bibinfo  {journal} {Phys.~Rev.~B}\ }\textbf {\bibinfo {volume} {63}},\
  \bibinfo {pages} {165314} (\bibinfo {year} {2001})}\BibitemShut {NoStop}%
\bibitem [{\citenamefont {Loss}\ and\ \citenamefont
  {Sukhorukov}(2000)}]{Loss2000}%
  \BibitemOpen
  \bibfield  {author} {\bibinfo {author} {\bibfnamefont {D.}~\bibnamefont
  {Loss}}\ and\ \bibinfo {author} {\bibfnamefont {E.~V.}\ \bibnamefont
  {Sukhorukov}},\ }\href {\doibase 10.1103/PhysRevLett.84.1035} {\bibfield
  {journal} {\bibinfo  {journal} {Phys.~Rev.~Lett.}\ }\textbf {\bibinfo
  {volume} {84}},\ \bibinfo {pages} {1035} (\bibinfo {year}
  {2000})}\BibitemShut {NoStop}%
\bibitem [{\citenamefont {Rosdahl}\ \emph {et~al.}(2018)\citenamefont
  {Rosdahl}, \citenamefont {Vuik}, \citenamefont {Kjaergaard},\ and\
  \citenamefont {Akhmerov}}]{Rosdahl2018}%
  \BibitemOpen
  \bibfield  {author} {\bibinfo {author} {\bibfnamefont {T.~O.}\ \bibnamefont
  {Rosdahl}}, \bibinfo {author} {\bibfnamefont {A.}~\bibnamefont {Vuik}},
  \bibinfo {author} {\bibfnamefont {M.}~\bibnamefont {Kjaergaard}}, \ and\
  \bibinfo {author} {\bibfnamefont {A.~R.}\ \bibnamefont {Akhmerov}},\ }\href
  {\doibase 10.1103/PhysRevB.97.045421} {\bibfield  {journal} {\bibinfo
  {journal} {Phys.~Rev.~B}\ }\textbf {\bibinfo {volume} {97}},\ \bibinfo
  {pages} {045421} (\bibinfo {year} {2018})}\BibitemShut {NoStop}%
\bibitem [{Men()}]{Menard2019}%
  \BibitemOpen
  \href@noop {} {}\bibinfo {note} {G. C. M\`{e}nard \textit{et al.},
  unpublished}\BibitemShut {NoStop}%
\bibitem [{\citenamefont {Takane}\ and\ \citenamefont
  {Ebisawa}(1992)}]{Takane1992}%
  \BibitemOpen
  \bibfield  {author} {\bibinfo {author} {\bibfnamefont {Y.}~\bibnamefont
  {Takane}}\ and\ \bibinfo {author} {\bibfnamefont {H.}~\bibnamefont
  {Ebisawa}},\ }\href {\doibase 10.1143/JPSJ.61.1685} {\bibfield  {journal}
  {\bibinfo  {journal} {J.~Phys.~Soc.~Jpn.}\ }\textbf {\bibinfo {volume}
  {61}},\ \bibinfo {pages} {1685} (\bibinfo {year} {1992})}\BibitemShut
  {NoStop}%
\bibitem [{\citenamefont {Aleiner}\ \emph {et~al.}(2002)\citenamefont
  {Aleiner}, \citenamefont {Brouwer},\ and\ \citenamefont
  {Glazman}}]{AleinerReview}%
  \BibitemOpen
  \bibfield  {author} {\bibinfo {author} {\bibfnamefont {I.}~\bibnamefont
  {Aleiner}}, \bibinfo {author} {\bibfnamefont {P.}~\bibnamefont {Brouwer}}, \
  and\ \bibinfo {author} {\bibfnamefont {L.}~\bibnamefont {Glazman}},\ }\href
  {\doibase 10.1016/S0370-1573(01)00063-1} {\bibfield  {journal} {\bibinfo
  {journal} {Physics Reports}\ }\textbf {\bibinfo {volume} {358}},\ \bibinfo
  {pages} {309} (\bibinfo {year} {2002})}\BibitemShut {NoStop}%
\bibitem [{\citenamefont {Hansen}\ \emph {et~al.}(2018)\citenamefont {Hansen},
  \citenamefont {Danon},\ and\ \citenamefont {Flensberg}}]{Hansen2018}%
  \BibitemOpen
  \bibfield  {author} {\bibinfo {author} {\bibfnamefont {E.~B.}\ \bibnamefont
  {Hansen}}, \bibinfo {author} {\bibfnamefont {J.}~\bibnamefont {Danon}}, \
  and\ \bibinfo {author} {\bibfnamefont {K.}~\bibnamefont {Flensberg}},\ }\href
  {\doibase 10.1103/PhysRevB.97.041411} {\bibfield  {journal} {\bibinfo
  {journal} {Phys.~Rev.~B}\ }\textbf {\bibinfo {volume} {97}},\ \bibinfo
  {pages} {041411(R)} (\bibinfo {year} {2018})}\BibitemShut {NoStop}%
\bibitem [{\citenamefont {Oreg}\ \emph {et~al.}(2010)\citenamefont {Oreg},
  \citenamefont {Refael},\ and\ \citenamefont {von Oppen}}]{Oreg2010}%
  \BibitemOpen
  \bibfield  {author} {\bibinfo {author} {\bibfnamefont {Y.}~\bibnamefont
  {Oreg}}, \bibinfo {author} {\bibfnamefont {G.}~\bibnamefont {Refael}}, \ and\
  \bibinfo {author} {\bibfnamefont {F.}~\bibnamefont {von Oppen}},\ }\href
  {\doibase 10.1103/PhysRevLett.105.177002} {\bibfield  {journal} {\bibinfo
  {journal} {Phys.~Rev.~Lett.}\ }\textbf {\bibinfo {volume} {105}},\ \bibinfo
  {pages} {177002} (\bibinfo {year} {2010})}\BibitemShut {NoStop}%
\bibitem [{\citenamefont {Lutchyn}\ \emph {et~al.}(2010)\citenamefont
  {Lutchyn}, \citenamefont {Sau},\ and\ \citenamefont
  {Das~Sarma}}]{Lutchyn2010}%
  \BibitemOpen
  \bibfield  {author} {\bibinfo {author} {\bibfnamefont {R.~M.}\ \bibnamefont
  {Lutchyn}}, \bibinfo {author} {\bibfnamefont {J.~D.}\ \bibnamefont {Sau}}, \
  and\ \bibinfo {author} {\bibfnamefont {S.}~\bibnamefont {Das~Sarma}},\ }\href
  {\doibase 10.1103/PhysRevLett.105.077001} {\bibfield  {journal} {\bibinfo
  {journal} {Phys. Rev. Lett.}\ }\textbf {\bibinfo {volume} {105}},\ \bibinfo
  {pages} {077001} (\bibinfo {year} {2010})}\BibitemShut {NoStop}%
\bibitem [{\citenamefont {Ben-Shach}\ \emph {et~al.}(2015)\citenamefont
  {Ben-Shach}, \citenamefont {Haim}, \citenamefont {Appelbaum}, \citenamefont
  {Oreg}, \citenamefont {Yacoby},\ and\ \citenamefont
  {Halperin}}]{Ben-Shach2015}%
  \BibitemOpen
  \bibfield  {author} {\bibinfo {author} {\bibfnamefont {G.}~\bibnamefont
  {Ben-Shach}}, \bibinfo {author} {\bibfnamefont {A.}~\bibnamefont {Haim}},
  \bibinfo {author} {\bibfnamefont {I.}~\bibnamefont {Appelbaum}}, \bibinfo
  {author} {\bibfnamefont {Y.}~\bibnamefont {Oreg}}, \bibinfo {author}
  {\bibfnamefont {A.}~\bibnamefont {Yacoby}}, \ and\ \bibinfo {author}
  {\bibfnamefont {B.~I.}\ \bibnamefont {Halperin}},\ }\href {\doibase
  10.1103/PhysRevB.91.045403} {\bibfield  {journal} {\bibinfo  {journal}
  {Phys.~Rev.~B}\ }\textbf {\bibinfo {volume} {91}},\ \bibinfo {pages} {045403}
  (\bibinfo {year} {2015})}\BibitemShut {NoStop}%
\end{thebibliography}
\end{document}